\documentclass[a4paper]{cas-sc}
\graphicspath{{fig/}}
\usepackage{lineno}
\linenumbers

\usepackage{lipsum}
\usepackage{hyperref}
\usepackage{booktabs}
\usepackage[numbers,sort&compress]{natbib}
\usepackage{nccmath}

\let\printorcid\relax % Remove ORCID footnote
\begin{document}

\title [mode = title]{Operation of BGO with SiPM readout at dry ice and liquid nitrogen temperatures}

\author[a]{K. Ding}
\author[a]{J. Liu}

\affiliation[a]{Department of Physics, University of South Dakota, 414 East Clark Street, Vermillion, SD 57069, USA}

\begin{abstract}
The light yield and decay constant of BGO were measured at both dry ice and liquid nitrogen temperatures using two SiPMs directly coupled to a $6\times6\times6$ cm$^2$ cubic BGO crystal. With the measured light yield (5.2$\pm$0.3 PE/keV at dry ice temperature and 10.5$\pm$0.4 PE/keV at liquid nitrogen temperature) and decay constants, potential applications of BGO in ToF-PET and SPECT were discussed.
\end{abstract}


\maketitle

\section{Introduction}
The Bismuth Germanate (Bi$_4$Ge$_3$O$_{12}$, BGO in short) scintillating crystal boasts a few properties that make it great for radiation detection in general. First, it is not hygroscopic, not fragile, hence is easy to handle. Second, it can be obtained at a reasonable price and is among the cost-effective solutions for many applications. Third, the high density (7.13~g/cm$^3$) and the effective atomic number (74) of BGO make it very effective in absorbing $\gamma$-ray radiation~\cite{scintillation_2000}, such as the 511~keV $\gamma$-ray utilized in positron emission tomography (PET)~\cite{schaart21, PET1, PET2}, the 141~keV $\gamma$-ray utilized in single photon emission computerized tomography (SPECT)~\cite{spect}, or Compton-scattered $\gamma$-rays from more sensitive detectors, such as high-purity germanium detectors~\cite{luminescence_1973, GeBGO}. Finally, BGO also finds its applications in fundamental science research. The high atomic number of Bi leads to its high cross-sections for neutrinos and dark matter particles coherent scatterings, which is a desirable property in detecting these rare physics phenomena~\cite{BGODM}.

However, the intrinsic light yield ($Y$) of BGO is not particularly high, only about 15--20\% of that of NaI(Tl). It is also a relatively slow scintillator, the decay time ($\tau_\text{decay}$) is about 300~ns at room temperature. $\sqrt{Y/\tau_\text{decay}}$ is a commonly used figure-of-merit (FOM) to assess how well a  scintillator can provide time information in modern PET that utilizes the time-of-flight (ToF) technique to improve image resolution~\cite{schaart21}. According to this FOM, Lutetium Oxyorthosilicate (Lu2SiO5:Ce, LSO in short) and Lutetium-Yttrium Oxyorthosilicate (LYSO) are about 10 times better than BGO in the application of ToF-PET~\cite{schaart21}, as they possess about three times higher light yields and about 10 times shorter decay times. However, L(Y)SO suffers from intrinsic radiation and is much more costly than BGO. If there is some way to modify the light yield and decay time of BGO, we may find a niche for BGO in medical imaging applications.

It has been observed that the light yield of BGO increases as the temperature goes down~\cite{BGO, BGO6K}. At liquid nitrogen temperature, its light yield is about five times higher than that at room temperature. Unfortunately, its decay time also increases as the temperature goes down~\cite{BGO6K}. At liquid nitrogen temperature, its decay constant is about thirty times longer than that at room temperature. It is hence of interest to find a temperature in between, where the light yield is already significantly higher than that at room temperature, but the decay time has not yet become too long. The dry ice temperature, which is convenient to obtain, seems to be a reasonable choice.

Cherenkov emission is much faster than scintillation. There is an increasing interest in the ToF-PET community to utilize it for better timing resolution (See Ref.~\cite{PET1} and references therein). A photo-electron created by a 511 keV gamma-ray in BGO can create on average $\sim20$ Cherenkov photons in the range of 305--750 nm in about 20 picoseconds (ps), about twice more and faster than that of LSO~\cite{PET1}. If one uses the Cherenkov emission in BGO for timing information and scintillation for energy information, the prolonged decay time of BGO after cooling becomes less a concern.

The readout of light signals at cryogenic temperatures necessitates the use of cryogenic light sensors. In recent years, cryogenic-compatible light sensors have emerged, with silicon photomultiplier (SiPM) being one of them. SiPMs offer several advantages over photomultiplier tubes (PMTs), including higher photon detection efficiency (PDE), absence of high voltage requirements, and reduced bulkiness. One drawback of SiPMs compared to PMTs is that they have a much higher dark count rate (DCR) than PMTs at room temperature. Fortunately, the rate drops quickly as the temperature goes down. Therefore, exploring the combination of BGO and SiPM at cryogenic temperatures is a worthwhile endeavor.

The proposed combination also possesses a few advantages over the traditional NaI(Tl) + PMT arrays in a typical SPECT system. First, as BGO is almost twice denser than NaI(Tl), it can be made twice thinner (about 0.3~cm) than the latter while maintaining the same absorption efficiency for 141~keV $\gamma$-rays. This will dramatically increase the camera's spatial resolution since less photons can be spread out to photon sensors far away~\cite{spect07, spect11, spect22, spect23}. Second, it possesses all advantages of SiPMs over PMTs as mentioned previously. The light yield of BGO at room temperature is only about 25\% of that of NaI(Tl). However, this may be overcome by the cryogenic operation of BGO.

This paper reports measurements of light yields and decay constants from a cubic BGO crystal directly coupled with two SensL SiPMs at both liquid nitrogen and dry ice temperatures. The 59.5 keV $\gamma$-ray peak from a $^{241}$Am source, the 30.85 keV $X$-ray peak, and the 81 keV $\gamma$-ray peak from a $^{133}$Ba source were used for the measurements. To the authors' knowledge, such an operation has not yet been reported in the literature.

\section{Experimental setup}
\label{s:exp}
\begin{figure}[htbp]
  \includegraphics[width=\linewidth]{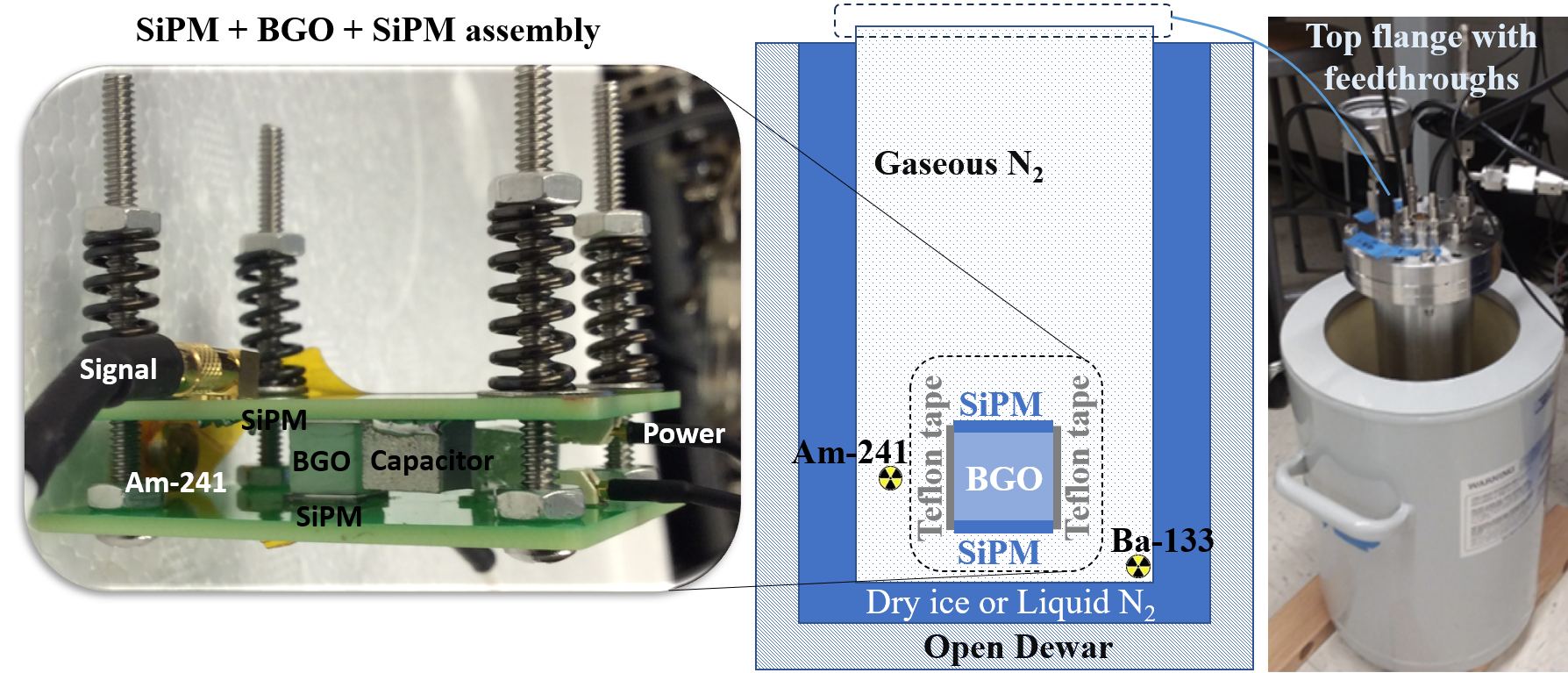}
  \caption{A picture and a sketch of experimental setups.}
  \label{f:setup}
\end{figure}

The experimental setup is shown in Fig.~\ref{f:setup}. The $6\times6\times6$~mm$^3$ cubic BGO crystal was purchased from \url{https://www.x-zlab.com}. To prevent any light leakage, the side surfaces of the crystal were wrapped with multiple layers of Teflon tape. The Teflon wrapping also randomizes the reflection direction to avoid trapping a photon in the crystal due to total internal reflection. The crystal surfaces were unpolished for the same reason. Two MicroFJ-SMTPA-60035 SiPMs from SensL~\cite{sipmJ} were directly coupled to the top and bottom surfaces of the crystal without optical grease, which may degrade at cryogenic temperature. A constant bias voltage of 29 V was applied to both sensors. The two sensors were directly soldered onto two passive boards. Detailed circuit diagrams and PCB layouts can be found in Ref.~\cite{ding22}. To establish optimal optical contact without the use of optical grease, we secured the PCBs against the crystal's end surfaces using springs. Two radioactive sources were used separately. An $^{241}$Am source was positioned by the side of the crystal, as shown in Fig.~\ref{f:setup}. A $^{133}$Ba source was placed at the bottom of the cryostat, slightly further away from the crystal compared to $^{241}$Am.

Given BGO's non-hygroscopic nature, setups were assembled in a standard room environment. The SiPM-BGO assembly was then lowered into a stainless steel chamber through its top opening, as illustrated in the schematic in the middle of Fig.~\ref{f:setup}. The chamber had an inner diameter of approximately 10 cm and a length of 50 cm. Both ends of the chamber were hermetically sealed under vacuum conditions, using two 6-inch ConFlat (CF) flanges. The bottom flange was blank and affixed to the chamber with a copper gasket in between. For ease of access during multiple operations, the top flange was attached to the chamber with a fluorocarbon CF gasket. The top flange featured vacuum-welded connections for five BNC, two SHV, one 19-pin electronic feedthrough, and two 1/4-inch VCR connectors.

Once all the cables were securely positioned within the chamber, the top flange was sealed. The chamber was subsequently evacuated using a Pfeiffer Vacuum HiCube 80 Eco pump to reach a vacuum level of approximately $1.0\times {10}^{-4}$~mbar. Subsequently, it was pressurized with dry nitrogen gas to approximately 1.8 Kgf/cm$^2$ and then carefully placed into an open LN$_2$ dewar. The dewar was subsequently filled with either LN$_2$ or dry ice to cool both the chamber and its contents. 

A few Heraeus C~220 platinum resistance temperature sensors were employed to monitor the cooling process effectively. These sensors were affixed to the side surface of the crystal, the bottom flange, and the top flange to obtain the temperature profile of the long chamber. For data acquisition from the sensors, we utilized a Raspberry Pi 2 computer running custom software~\cite{cravis}. The cooling process took about half an hour due to the small size of the crystal. Most measurements, however, were conducted after approximately two hours of waiting to ensure the system reached thermal equilibrium. Notably, the temperature of the crystal remained consistent before and after each measurement.

The passive boards were powered by a RIGOL DP821A DC power supply~\cite{dp800}. A voltage of 29~V was applied to the SiPMs. According to their manuals, the photon detection efficiency (PDE) at this voltage is $\sim 50\%$ for MicroFJ-SMTPA-60035 at 420~nm and room temperature. Signals were further amplified using a Phillips Scientific Quad Bipolar Amplifier Model 771, featuring four channels, each with a gain of ten. By chaining two channels together, a maximum gain of 10$\times$10 was employed. The amplified pulses were subsequently routed into a CAEN DT5720 waveform digitizer, with a 250~MHz sampling rate, a dynamic range of 2~V, and a 12-bit resolution. Data recording was carried out using WaveDump~\cite{wavedump}, a free software tool provided by CAEN. The recorded binary data files were later converted to CERN ROOT files for analysis~\cite{towards}. 

\section{Single photo-electron response}
\begin{figure}[htbp]\centering
  \includegraphics[width=0.49\linewidth]{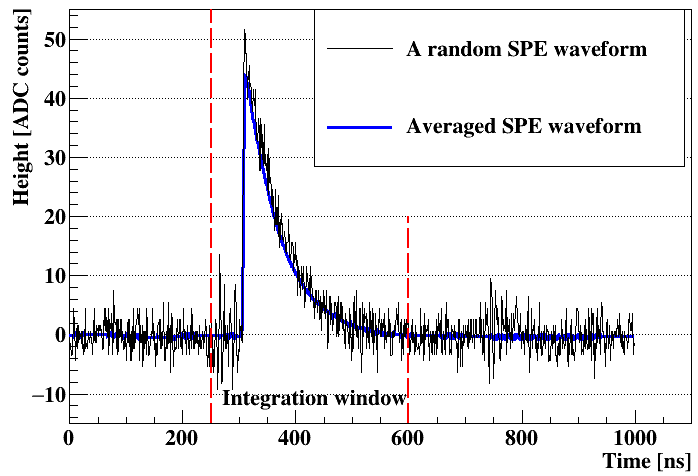}
  \includegraphics[width=0.49\linewidth]{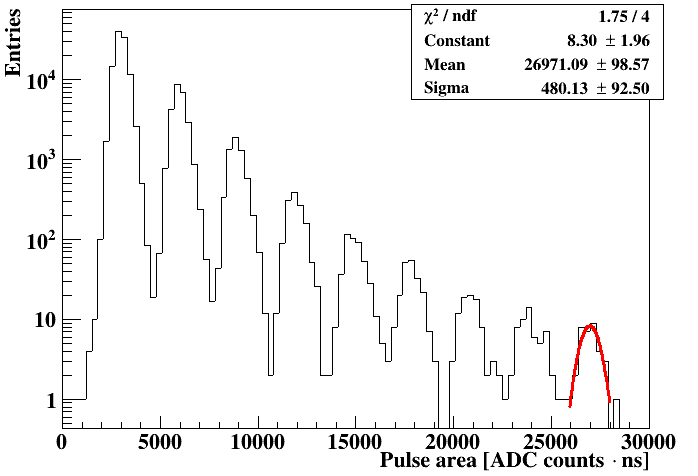}
  \caption{Left: A random single photon-electron (SPE) waveform and the averaged SPE waveform meaasured with the top SiPM at 195~K. Right: SPE distribution in logarithm scale with a Gaussian fitted to the last peak. Note: SPE responses from other temperatures and the other SiPM are very similar.}
  \label{f:single}
\end{figure}

Single photo-electron (SPE) responses of individual channels were investigated by analyzing waveform data triggered by dark counts. To filter out noisy events while still capturing the majority of SPE pulses, a threshold of 20 ADC counts was set. Prior to the rising edge of the pulse that triggered the digitizer, some pre-traces (100 samples) were preserved to calculate the baseline value of a waveform. This baseline value was subsequently subtracted from each sample of the waveform. The left plot in Fig.~\ref{f:single} shows a random single PE waveform overlaid with the averaged SPE waveform. The averaged SPE waveform was generated by adding 1000 SPE pulses together and then dividing the resulting summed waveform by 1000. Random electronic noise was canceled out in the averaged SPE waveform and a stable baseline was revealed. The integration window, [250, 600]~ns, was determined based on the averaged SPE waveform.

The right plot in Fig.~\ref{f:single} shows the distribution of integrated pulse areas. Individual photo-electron (PE) peaks are clearly distinguishable. The ninth peak was fitted using a Gaussian function to obtain its mean value. The same operations were performed for all other peaks. The mean of an SPE, mean$_\text{1PE}$, is defined as the Gaussian mean divided by the number of PEs, $n$. For instance, the mean$_\text{SPE}$ calculated using the ninth peak is $26971 / 9=2996$ ADC counts$\cdot$ns.  The discrepancy of mean$_\text{SPE}$ from different peaks is within 3\%. The same measurement was repeated multiple times before and after an energy calibration measurement. The discrepancy of mean$_\text{SPE}$ in multiple measurements was within 6\%. which was taken as the uncertainty of mean$_\text{SPE}$. 

To summarize, means$_\text{SPE}$ obtained from the top SiPM in the dry ice temperature runs is 2996 ADC counts$\cdot$ns, and for the bottom SiPM, it is 2972 ADC counts$\cdot$ns. Provided the same voltage, the gain of SiPM increases when the operating temperature decreases. Means$_\text{SPE}$ from the top SiPM in the liquid nitrogen temperature runs increased to 3716 ADC counts$\cdot$ns, and for the bottom SiPM, it is 3713 ADC counts$\cdot$ns.

\section{Energy calibration}
\label{s:ec}
The energy calibration was performed using $X$-rays and $\gamma$-rays from the $^{241}$Am and the $^{133}$Ba radioactive sources individually. The digitizer was triggered when the heights of pulses from both SiPMs were more than 100 ADC counts ($\sim$2 to 3 PEs). Pulses induced by radiation from the source were well above the threshold. The coincident trigger rate was around 240~Hz at dry ice temperature.

\begin{figure}[htbp]\centering
  \includegraphics[width=0.5\linewidth]{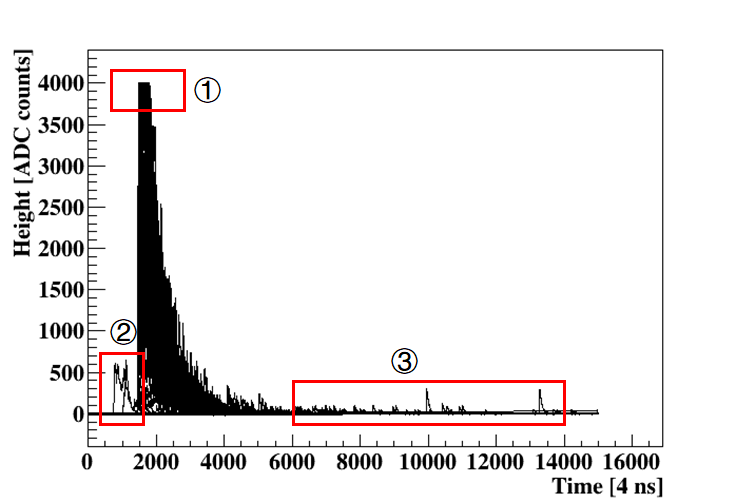}
  \caption{One hundred random waveforms from the top SiPM at dry ice temperature.}
  \label{f:wfs}
\end{figure}

Several waveform selection criteria were applied. First, the room mean square (RMS) of the first 100 baseline samples was calculated for each waveform. If it was more than 3.5 ADC counts, that waveform was rejected. Second, waveforms that went beyond the digitizer's dynamic range (region \textcircled{\scriptsize 1} in Fig.~\ref{f:wfs}) were rejected.  Third, waveforms that had pre-pulses above the threshold (region \textcircled{\scriptsize 2} in fig.~\ref{f:wfs}) was removed. Only waveforms triggered within [1450, 1460]~ns were selected. 

\begin{figure}[htbp]\centering
  \includegraphics[width=0.45\linewidth]{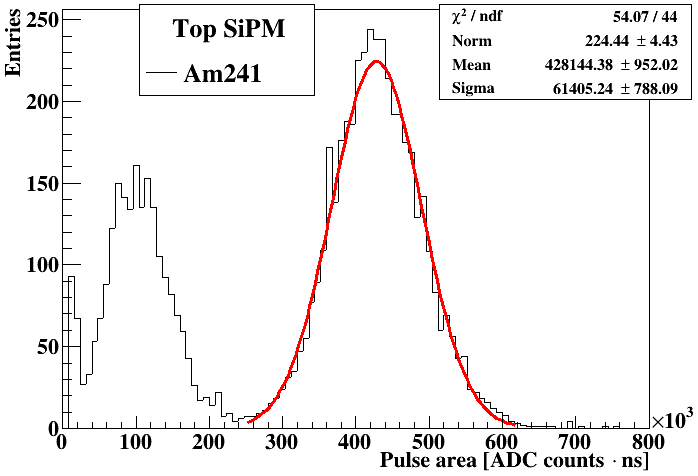}
  \includegraphics[width=0.45\linewidth]{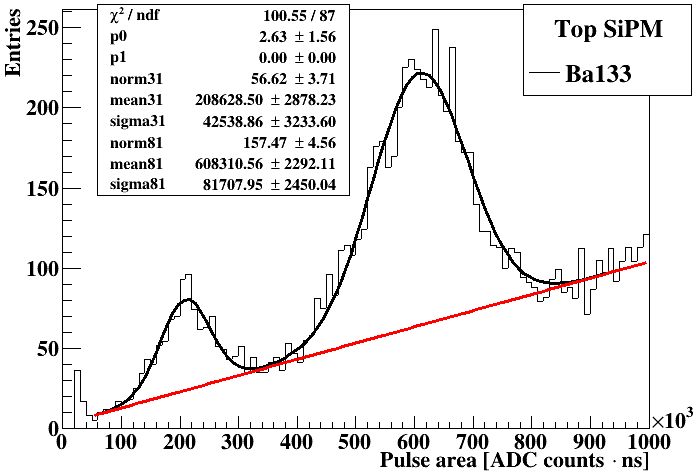}
  \caption{Energy spectra of $^{241}$Am (left) and $^{133}$Ba (right) observed from the top SiPM at dry ice temperature. Gaussian fitting were performed for three peaks (from left to right: 59.5 keV from $^{241}$Am, 30.85 and 81 keV from $^{133}$Ba).}
  \label{f:ec}
\end{figure}

At dry ice temperature, integration in [1400, 6000] ns was performed for waveforms passing all criteria. The integration has a unit of ADC counts$\cdot$ns. The resulting energy spectra of $^{241}$Am and $^{133}$Ba are presented in Fig.~\ref{f:ec}. It seems from Fig.~\ref{f:wfs} that tails of some pulses go beyond 6000 ns into region \textcircled{\scriptsize 3}. However, the averaged waveform calculated using the 81~keV peak (right most peak in Fig.~\ref{f:ec}) shown in the left plot in Fig.~\ref{f:tau} in Section~\ref{s:dc} justifies the range of integration. At liquid nitrogen temperature, light pulses were significantly prolonged as shown in the right plot in Fig.~\ref{f:tau}. The corresponding integration range is [1400, 12000] ns.

\section{Light yield}
\label{s:ly}
The integrated pulse areas, $A$, of the selected waveforms follow the Gaussian distribution as shown in Fig.~\ref{f:ec}. In the left plot, a Gaussian was fitted to the 59.5 keV peak from $^{241}$Am. In the right plot, a straight line was used to characterize the background, while two combined Gaussian functions were used to fit the 30.85 keV and 81 keV peaks from $^{133}$Ba. The fitting results were summarized in Table~\ref{t:dit}--\ref{t:lnb}.

The fitted means of $A$ corresponding to the 30.85 keV, 59.5 keV and 81 keV peaks in the $^{241}$Am and $^{133}$Ba spectra in the unit of ADC counts$\cdot$ns were converted to the number of PE using the following formula:
\begin{equation}
  \text{number of PEs} = \frac{\text{Mean  (A) [ADC counts} \cdot \text{ns]}}{\text{mean}_\text{SPE}},
  \label{e:m1pe}
\end{equation}
and the light yield was calculated using the data in Table~\ref{t:dit}--\ref{t:lnb} and the following formula:
\begin{equation}
  \text{light yield }[\text{PE/keV}_\text{ee}] = \frac{\text{number of PEs}}{\text{Energy }[\text{keV}_\text{ee}]}.
  \label{e:ly}
\end{equation}

Combining the results from both the top and the bottom SiPM, a light yield of 5.2 $\pm$ 0.3 PE/keV$_\text{ee}$ at 81 keV was observed at dry ice temperature, and 10.5 $\pm$ 0.4 PE/keV$_\text{ee}$ was observed at liquid nitrogen temperature. The yield decreases slightly as the energy goes down. This could be explained by either the observed non-linear energy response of BGO~\cite{LYbgo1, LYbgo2}, or the fact that lower energy photons cannot penetrate deep into the crystal, they produce scintillation light much closer to the rough surface of the crystal, and the produced scintillation light has higher chance to be trapped or absorbed by surface defects.

\begin{table}[htbp]\centering
  \caption{\label{t:dit} Light yield obtained at dry ice temperature from the top SiPM.}
  \begin{tabular}{l @{\extracolsep{\fill}}ccccccc}
    \toprule
    Type of & Source & Energy & Mean (A) & Sigma & FWHM & Light yield & Uncertainty\\
   radiation &  & [keV] & [ADC$\cdot$ns] & [ADC$\cdot$ns] & \% & [PE/keV$_\text{ee}$] & [PE/keV$_\text{ee}$] \\
    \midrule
     \begin{tabular}{r} $\gamma$-ray \\ $\gamma$-ray \\ $\gamma$-ray \\
     \end{tabular} &
     \begin{tabular}{r} Ba-133 \\ Am-241 \\ Ba-133 \\
     \end{tabular} &
     \begin{tabular}{l} 30.85 \\ 59.5 \\ 81 \\
     \end{tabular} &
     \begin{tabular}{r} 208629 \\ 428144 \\ 608311\\
     \end{tabular} &
     \begin{tabular}{r} 42539 \\ 61405 \\ 81708 \\
     \end{tabular} &
     \begin{tabular}{r} 48 \\ 34 \\ 32 \\
     \end{tabular} &
     \begin{tabular}{r} 2.3 \\ 2.4 \\ 2.5 \\
     \end{tabular} &
    \begin{tabular}{r} $\pm$ 0.1 \\ $\pm$ 0.1 \\ $\pm$ 0.2 \\
     \end{tabular} \\\bottomrule
     \end{tabular}
\end{table}

\begin{table}[htbp]\centering
  \caption{\label{t:dib} Light yield obtained at dry ice temperature from the bottom SiPM.}
  \begin{tabular}{l @{\extracolsep{\fill}}ccccccc}
    \toprule
    Type of & Source & Energy & Mean (A) & Sigma & FWHM & Light yield & Uncertainty\\
   radiation &  & [keV] & [ADC$\cdot$ns] & [ADC$\cdot$ns] & \% & [PE/keV$_\text{ee}$] & [PE/keV$_\text{ee}$] \\
    \midrule
     \begin{tabular}{r} $\gamma$-ray \\ $\gamma$-ray \\ $\gamma$-ray \\
     \end{tabular} &
     \begin{tabular}{r} Ba-133 \\ Am-241 \\ Ba-133 \\
     \end{tabular} &
     \begin{tabular}{l} 30.85 \\ 59.5 \\ 81 \\
     \end{tabular} &
     \begin{tabular}{r} 231940 \\ 427219 \\ 657506\\
     \end{tabular} &
     \begin{tabular}{r} 45189 \\ 62083 \\ 91739 \\
     \end{tabular} &
     \begin{tabular}{r} 46 \\ 34 \\ 33 \\
     \end{tabular} &
     \begin{tabular}{r} 2.5 \\ 2.4 \\ 2.7 \\
     \end{tabular} &
    \begin{tabular}{r} $\pm$ 0.1 \\ $\pm$ 0.1 \\ $\pm$ 0.2 \\
     \end{tabular} \\\bottomrule
     \end{tabular}
\end{table}

\begin{table}[htbp]\centering
  \caption{\label{t:lnt} Light yield obtained at liquid nitrogen temperature from the top SiPM.}
  \begin{tabular}{l @{\extracolsep{\fill}}ccccccc}
    \toprule
    Type of & Source & Energy & Mean (A) & Sigma & FWHM & Light yield & Uncertainty\\
   radiation &  & [keV] & [ADC$\cdot$ns] & [ADC$\cdot$ns] & \% & [PE/keV$_\text{ee}$] & [PE/keV$_\text{ee}$] \\
    \midrule
     \begin{tabular}{r} $\gamma$-ray \\ $\gamma$-ray \\ $\gamma$-ray \\
     \end{tabular} &
     \begin{tabular}{r} Ba-133 \\ Am-241 \\ Ba-133 \\
     \end{tabular} &
     \begin{tabular}{l} 30.85 \\ 59.5 \\ 81 \\
     \end{tabular} &
     \begin{tabular}{r} 477785 \\ 1095347 \\ 1515401\\
     \end{tabular} &
     \begin{tabular}{r} 131485 \\ 166360 \\ 195567 \\
     \end{tabular} &
     \begin{tabular}{r} 65 \\ 36 \\ 30 \\
     \end{tabular} &
     \begin{tabular}{r} 4.2 \\ 5.0 \\ 5.0 \\
     \end{tabular} &
    \begin{tabular}{r} $\pm$ 0.2 \\ $\pm$ 0.3 \\ $\pm$ 0.3 \\
     \end{tabular} \\\bottomrule
     \end{tabular}
\end{table}

\begin{table}[htbp]\centering
  \caption{\label{t:lnb} Light yield obtained at liquid nitrogen temperature from the bottom SiPM.}
  \begin{tabular}{l @{\extracolsep{\fill}}ccccccc}
    \toprule
    Type of & Source & Energy & Mean (A) & Sigma & FWHM & Light yield & Uncertainty\\
   radiation &  & [keV] & [ADC$\cdot$ns] & [ADC$\cdot$ns] & \% & [PE/keV$_\text{ee}$] & [PE/keV$_\text{ee}$] \\
    \midrule
     \begin{tabular}{r} $\gamma$-ray \\ $\gamma$-ray \\ $\gamma$-ray \\
     \end{tabular} &
     \begin{tabular}{r} Ba-133 \\ Am-241 \\ Ba-133 \\
     \end{tabular} &
     \begin{tabular}{l} 30.85 \\ 59.5 \\ 81 \\
     \end{tabular} &
     \begin{tabular}{r} 549115 \\ 1166037 \\ 1655799\\
     \end{tabular} &
     \begin{tabular}{r} 117970 \\ 173975 \\ 222286 \\
     \end{tabular} &
     \begin{tabular}{r} 51 \\ 35 \\ 32 \\
     \end{tabular} &
     \begin{tabular}{r} 4.8 \\ 5.3 \\ 5.5 \\
     \end{tabular} &
    \begin{tabular}{r} $\pm$ 0.3 \\ $\pm$ 0.3 \\ $\pm$ 0.3 \\
     \end{tabular} \\\bottomrule
     \end{tabular}
\end{table}

\section{Decay constant}
\label{s:dc}
\begin{figure}[htbp]
  \includegraphics[width=0.49\linewidth]{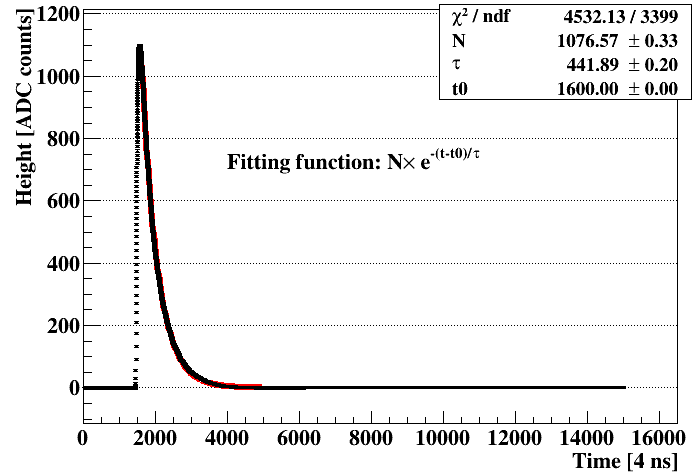}
  \includegraphics[width=0.49\linewidth]{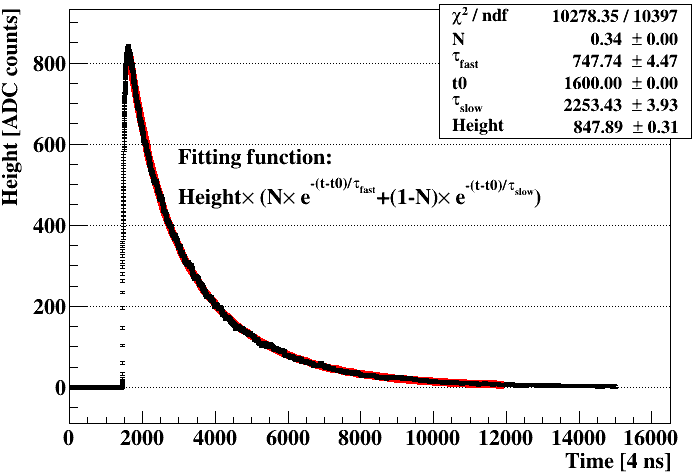}
  \caption{Averaged 81 keV waveforms at dry ice temperature (left) and liquid nitrogen temperature (right).}
  \label{f:tau}
\end{figure}

The decay constants of BGO at dry ice (195~K) and liquid nitrogen (77~K) temperatures were obtained by fitting exponential functions to the averaged waveforms in the 81 keV peak. The process of generating an averaged waveform is explained using the data from the top SiPM at 195~K. First, waveforms that have a pulse area in [550,000, 650,000] ADC counts$\cdot$ns were selected (less than $\pm1\sigma$ around the mean of the 81 keV peak shown in Fig.~\ref{f:ec}). Second, the selected waveforms were summed. Third, the summed waveform was divided by the total number of selected waveforms to obtained the averaged waveform, which is shown in the left plot in Fig.~\ref{f:tau}. The averaged waveform at liquid nitrogen temperature is shown in the right plot.

The region selected for fitting is [1600, 6000]$]\times$4 ns, while at 77~K is [1600, 12000]$\times$4 ns. The corresponding fitting functions and results are shown in Fig.~\ref{f:tau}. A single decay constant, $442 \times 4 = 1768$ ns, was observed at 195~K. A fast decay constant, $748 \times 4 = 2992$ ns, and a slow decay constant, $2253 \times 4 = 9012$ ns, were observed at 77~K. The observed upward trend in the decay constant with the decreasing temperature in BGO is consistent with the literature~\cite{BGO6K}.

\section{Discussion}
Using the measured light yields and decay constants, FOM of BGO for ToF-PET can be calculated. Note that the light yield defined in Section~\ref{s:ly} is the yield of the whole detector system, not the intrinsic yield of the crystal. Since the latter is used to calculate the FOM in the literature~\cite{schaart21}, the following equation was used to convert the system yield to the intrinsic yield for easy comparison:
\begin{equation}
    \text{intrinsic light yield} = \text{(system light yield)/(PDE of SiPM)/(light collection efficiency)}
\end{equation}

According to the SiPM manual from the manufacturer~\cite{sensl}, the PDE at 480~nm (peak scintillation wavelengh of BGO) is about 45\%. The light collection efficiency from our Geant4 optical simulation is about 80\%. The intrinsic light yield is then 5.2/45\%/80\%$\sim$14 photons/keV (or 14,000 photons/MeV) at dry ice temperature, and 10.3/45\%/80\%$\sim$29 photons/keV (or 29,000 photons/MeV) at liquid nitrogen temperature. The FOM is then $\sqrt{14000/1768}=2.8/\sqrt{\text{MeV}\cdot\text{ns}}$ at dry ice temperature, and $\sqrt{29000/2992}=3.1/\sqrt{\text{MeV}\cdot\text{ns}}$  (fast decay constant), $\sqrt{29000/9012}=1.8/\sqrt{\text{MeV}\cdot\text{ns}}$ (slow decay constant) at liquid nitrogen temperature. They are about 10 times smaller than the typical FOM for LYSO/LSO crystals (27--32~$/\sqrt{\text{MeV}\cdot\text{ns}}$~\cite{schaart21}). The light yield of cooled BGO is compatible to that of LYSO/LSO. However, the significantly prolonged decay constant of cooled BGO cancels out all the benefit from the increased light yield.

One potential solution is to use the Cherenkov radiation emitted by photo-electrons created by incident $\gamma$-rays to obtain ToF and to use the scintillation light for energy information. The Cherenkov radiation in BGO is emitted within 30~ps~\cite{PET1}. The FOM calculated using this number is $\sqrt{29000/0.03}=983/\sqrt{\text{MeV}\cdot\text{ns}}$, more than 30 times better than that of LYSO/LSO.

The combination of cooled BGO + SiPM light readout also exhibits some great properties as the Anger Camera for a SPECT system, such as a light yield that is compatible to that of NaI(Tl) and a better spatial resolution, etc. The prolonged decay time is less a concern in a SPECT system as the time information is not utilized there.

\section{Conclusion}
The light yield and decay constant of BGO were measured at both dry ice and liquid nitrogen temperatures using two SiPMs directly coupled to a $6\times6\times6$ cm$^2$ cubic BGO crystal. With the measured light yield (5.2$\pm$0.3 PE/keV at dry ice temperature and 10.5$\pm$0.4 PE/keV at liquid nitrogen temperature) and decay constants, potential applications of BGO in ToF-PET and SPECT were discussed. The increased light yield is very beneficial for both applications. The prolonged decay time is less a concern in SPECT and may be overcome by using Cheronkov radiation instead in a ToF-PET system.

\section*{Acknowledgements}
This work is supported by the National Science Foundation (NSF), USA, award PHY-1506036, and the Grant-in-Aid for Encouragement of Young Scientists (B), No. 26800122, MEXT, Japan. Computations supporting this project were performed on High Performance Computing systems at the University of South Dakota, funded by NSF award OAC-1626516.

\bibliographystyle{achemso}
\bibliography{ref}

\providecommand{\latin}[1]{#1}
\makeatletter
\providecommand{\doi}
  {\begingroup\let\do\@makeother\dospecials
  \catcode`\{=1 \catcode`\}=2 \doi@aux}
\providecommand{\doi@aux}[1]{\endgroup\texttt{#1}}
\makeatother
\providecommand*\mcitethebibliography{\thebibliography}
\csname @ifundefined\endcsname{endmcitethebibliography}  {\let\endmcitethebibliography\endthebibliography}{}
\begin{mcitethebibliography}{23}
\providecommand*\natexlab[1]{#1}
\providecommand*\mciteSetBstSublistMode[1]{}
\providecommand*\mciteSetBstMaxWidthForm[2]{}
\providecommand*\mciteBstWouldAddEndPuncttrue
  {\def\EndOfBibitem{\unskip.}}
\providecommand*\mciteBstWouldAddEndPunctfalse
  {\let\EndOfBibitem\relax}
\providecommand*\mciteSetBstMidEndSepPunct[3]{}
\providecommand*\mciteSetBstSublistLabelBeginEnd[3]{}
\providecommand*\EndOfBibitem{}
\mciteSetBstSublistMode{f}
\mciteSetBstMaxWidthForm{subitem}{(\alph{mcitesubitemcount})}
\mciteSetBstSublistLabelBeginEnd
  {\mcitemaxwidthsubitemform\space}
  {\relax}
  {\relax}

\bibitem[Melcher(2000)]{scintillation_2000}
Melcher,~C.~L. Scintillation {Crystals} for {PET}. \emph{Journal of Nuclear Medicine} \textbf{2000}, \emph{41}, 1051--1055, Publisher: Society of Nuclear Medicine Section: Continuing Education\relax
\mciteBstWouldAddEndPuncttrue
\mciteSetBstMidEndSepPunct{\mcitedefaultmidpunct}
{\mcitedefaultendpunct}{\mcitedefaultseppunct}\relax
\EndOfBibitem
\bibitem[Schaart(2021)]{schaart21}
Schaart,~D.~R. Physics and technology of time-of-flight {PET} detectors. \emph{Physics in Medicine \& Biology} \textbf{2021}, \emph{66}, 09TR01\relax
\mciteBstWouldAddEndPuncttrue
\mciteSetBstMidEndSepPunct{\mcitedefaultmidpunct}
{\mcitedefaultendpunct}{\mcitedefaultseppunct}\relax
\EndOfBibitem
\bibitem[Brunner and Schaart(2017)Brunner, and Schaart]{PET1}
Brunner,~S.~E.; Schaart,~D.~R. {BGO} as a hybrid scintillator / {Cherenkov} radiator for cost-effective time-of-flight {PET}. \emph{Physics in Medicine \& Biology} \textbf{2017}, \emph{62}, 4421\relax
\mciteBstWouldAddEndPuncttrue
\mciteSetBstMidEndSepPunct{\mcitedefaultmidpunct}
{\mcitedefaultendpunct}{\mcitedefaultseppunct}\relax
\EndOfBibitem
\bibitem[Kwon \latin{et~al.}(2016)Kwon, Gola, Ferri, Piemonte, and Cherry]{PET2}
Kwon,~S.~I.; Gola,~A.; Ferri,~A.; Piemonte,~C.; Cherry,~S.~R. Bismuth germanate coupled to near ultraviolet silicon photomultipliers for time-of-flight {PET}. \emph{Physics in Medicine \& Biology} \textbf{2016}, \emph{61}, L38\relax
\mciteBstWouldAddEndPuncttrue
\mciteSetBstMidEndSepPunct{\mcitedefaultmidpunct}
{\mcitedefaultendpunct}{\mcitedefaultseppunct}\relax
\EndOfBibitem
\bibitem[Peterson and Furenlid(2011)Peterson, and Furenlid]{spect}
Peterson,~T.~E.; Furenlid,~L.~R. {SPECT} detectors: the {Anger} {Camera} and beyond. \emph{Physics in medicine and biology} \textbf{2011}, \emph{56}, R145--R182\relax
\mciteBstWouldAddEndPuncttrue
\mciteSetBstMidEndSepPunct{\mcitedefaultmidpunct}
{\mcitedefaultendpunct}{\mcitedefaultseppunct}\relax
\EndOfBibitem
\bibitem[Weber and Monchamp(1973)Weber, and Monchamp]{luminescence_1973}
Weber,~M.~J.; Monchamp,~R.~R. Luminescence of {Bi4} {Ge3} {O12} : {Spectral} and decay properties. \emph{Journal of Applied Physics} \textbf{1973}, \emph{44}, 5495--5499\relax
\mciteBstWouldAddEndPuncttrue
\mciteSetBstMidEndSepPunct{\mcitedefaultmidpunct}
{\mcitedefaultendpunct}{\mcitedefaultseppunct}\relax
\EndOfBibitem
\bibitem[de~Voigt \latin{et~al.}(1995)de~Voigt, Bacelar, Micek, Schotanus, Verhoef, Wintraecken, and Vermeulen]{GeBGO}
de~Voigt,~M. J.~A.; Bacelar,~J.~C.; Micek,~S.~L.; Schotanus,~P.; Verhoef,~B. A.~W.; Wintraecken,~Y. J.~E.; Vermeulen,~P. A novel compact {Ge}-{BGO} {Compton}-suppression spectrometer. \emph{Nuclear Instruments and Methods in Physics Research Section A: Accelerators, Spectrometers, Detectors and Associated Equipment} \textbf{1995}, \emph{356}, 362--375\relax
\mciteBstWouldAddEndPuncttrue
\mciteSetBstMidEndSepPunct{\mcitedefaultmidpunct}
{\mcitedefaultendpunct}{\mcitedefaultseppunct}\relax
\EndOfBibitem
\bibitem[Ortigoza()]{BGODM}
Ortigoza,~Y. {BGO} {Scintillating} {Bolometer}: {Its} application in dark matter experiments. \url{http://taup2009.lngs.infn.it/slides/jul3/ortigoza.pdf}\relax
\mciteBstWouldAddEndPuncttrue
\mciteSetBstMidEndSepPunct{\mcitedefaultmidpunct}
{\mcitedefaultendpunct}{\mcitedefaultseppunct}\relax
\EndOfBibitem
\bibitem[BGO()]{BGO}
{BGO} {\textbar} {Scintillation} {Crystal}. \url{https://www.x-zlab.com/product/bgo-scintillation-crystal/}\relax
\mciteBstWouldAddEndPuncttrue
\mciteSetBstMidEndSepPunct{\mcitedefaultmidpunct}
{\mcitedefaultendpunct}{\mcitedefaultseppunct}\relax
\EndOfBibitem
\bibitem[Gironnet(2008)]{BGO6K}
Gironnet,~J. Scintillation studies of {Bi4Ge3O12} ({BGO}) down to a temperature of 6 {K}. \emph{Nuclear Instruments and Methods in Physics Research Section A: Accelerators, Spectrometers, Detectors and Associated Equipment} \textbf{2008}, \emph{594}, 358--361\relax
\mciteBstWouldAddEndPuncttrue
\mciteSetBstMidEndSepPunct{\mcitedefaultmidpunct}
{\mcitedefaultendpunct}{\mcitedefaultseppunct}\relax
\EndOfBibitem
\bibitem[Madsen(2007)]{spect07}
Madsen,~M.~T. Recent {Advances} in {SPECT} {Imaging}. \emph{Journal of Nuclear Medicine} \textbf{2007}, \emph{48}, 661--673\relax
\mciteBstWouldAddEndPuncttrue
\mciteSetBstMidEndSepPunct{\mcitedefaultmidpunct}
{\mcitedefaultendpunct}{\mcitedefaultseppunct}\relax
\EndOfBibitem
\bibitem[Peterson and Furenlid(2011)Peterson, and Furenlid]{spect11}
Peterson,~T.~E.; Furenlid,~L.~R. {SPECT} detectors: the {Anger} {Camera} and beyond. \emph{Phys. Med. Biol.} \textbf{2011}, \emph{56}, R145--R182\relax
\mciteBstWouldAddEndPuncttrue
\mciteSetBstMidEndSepPunct{\mcitedefaultmidpunct}
{\mcitedefaultendpunct}{\mcitedefaultseppunct}\relax
\EndOfBibitem
\bibitem[Ritt(2022)]{spect22}
Ritt,~P. Recent {Developments} in {SPECT}/{CT}. \emph{Seminars in Nuclear Medicine} \textbf{2022}, \emph{52}, 276--285\relax
\mciteBstWouldAddEndPuncttrue
\mciteSetBstMidEndSepPunct{\mcitedefaultmidpunct}
{\mcitedefaultendpunct}{\mcitedefaultseppunct}\relax
\EndOfBibitem
\bibitem[Massari and Mok(2023)Massari, and Mok]{spect23}
Massari,~R.; Mok,~G. S.~P. Editorial: {New} trends in single photon emission computed tomography ({SPECT}). \emph{Front. Med.} \textbf{2023}, \emph{10}\relax
\mciteBstWouldAddEndPuncttrue
\mciteSetBstMidEndSepPunct{\mcitedefaultmidpunct}
{\mcitedefaultendpunct}{\mcitedefaultseppunct}\relax
\EndOfBibitem
\bibitem[sip()]{sipmJ}
{MicroFJ}-{SMTPA}-60035. \url{https://www.onsemi.com/support/evaluation-board/microfj-sma-60035-gevb}\relax
\mciteBstWouldAddEndPuncttrue
\mciteSetBstMidEndSepPunct{\mcitedefaultmidpunct}
{\mcitedefaultendpunct}{\mcitedefaultseppunct}\relax
\EndOfBibitem
\bibitem[Ding \latin{et~al.}(2022)Ding, Liu, Yang, and Chernyak]{ding22}
Ding,~K.; Liu,~J.; Yang,~Y.; Chernyak,~D. First operation of undoped {CsI} directly coupled with {SiPMs} at 77 {K}. \emph{Eur. Phys. J. C} \textbf{2022}, \emph{82}, 344\relax
\mciteBstWouldAddEndPuncttrue
\mciteSetBstMidEndSepPunct{\mcitedefaultmidpunct}
{\mcitedefaultendpunct}{\mcitedefaultseppunct}\relax
\EndOfBibitem
\bibitem[Mammo(2018)]{cravis}
Mammo,~J. Josephss/{CraViS}. 2018; \url{https://github.com/Josephss/CraViS}\relax
\mciteBstWouldAddEndPuncttrue
\mciteSetBstMidEndSepPunct{\mcitedefaultmidpunct}
{\mcitedefaultendpunct}{\mcitedefaultseppunct}\relax
\EndOfBibitem
\bibitem[dp8()]{dp800}
{DP800} {High} {Performance} {Linear} {DC} {Power} {Supplies} {\textbar} {RIGOL}. \url{https://www.rigolna.com/products/dc-power-loads/dp800/}\relax
\mciteBstWouldAddEndPuncttrue
\mciteSetBstMidEndSepPunct{\mcitedefaultmidpunct}
{\mcitedefaultendpunct}{\mcitedefaultseppunct}\relax
\EndOfBibitem
\bibitem[wav()]{wavedump}
{WaveDump} - {CAEN} {Digitizer} readout application. \url{https://www.caen.it/products/caen-wavedump/}\relax
\mciteBstWouldAddEndPuncttrue
\mciteSetBstMidEndSepPunct{\mcitedefaultmidpunct}
{\mcitedefaultendpunct}{\mcitedefaultseppunct}\relax
\EndOfBibitem
\bibitem[Liu(2021)]{towards}
Liu,~J. jintonic/toward. 2021; \url{https://github.com/jintonic/toward}\relax
\mciteBstWouldAddEndPuncttrue
\mciteSetBstMidEndSepPunct{\mcitedefaultmidpunct}
{\mcitedefaultendpunct}{\mcitedefaultseppunct}\relax
\EndOfBibitem
\bibitem[Yawai \latin{et~al.}(2014)Yawai, Chewpraditkul, Wanarak, Nikl, and Ratanatongchai]{LYbgo1}
Yawai,~N.; Chewpraditkul,~W.; Wanarak,~C.; Nikl,~M.; Ratanatongchai,~W. Intrinsic light yield and light loss coefficient of {Bi4Ge3O12} single crystals. \emph{Optical Materials} \textbf{2014}, \emph{36}, 2030--2033\relax
\mciteBstWouldAddEndPuncttrue
\mciteSetBstMidEndSepPunct{\mcitedefaultmidpunct}
{\mcitedefaultendpunct}{\mcitedefaultseppunct}\relax
\EndOfBibitem
\bibitem[Moszynski \latin{et~al.}(2004)Moszynski, Balcerzyk, Czarnacki, Kapusta, Klamra, Syntfeld, and Szawlowski]{LYbgo2}
Moszynski,~M.; Balcerzyk,~M.; Czarnacki,~W.; Kapusta,~M.; Klamra,~W.; Syntfeld,~A.; Szawlowski,~M. Intrinsic energy resolution and light yield nonproportionality of {BGO}. \emph{IEEE Transactions on Nuclear Science} \textbf{2004}, \emph{51}, 1074--1079\relax
\mciteBstWouldAddEndPuncttrue
\mciteSetBstMidEndSepPunct{\mcitedefaultmidpunct}
{\mcitedefaultendpunct}{\mcitedefaultseppunct}\relax
\EndOfBibitem
\bibitem[sen()]{sensl}
Onsemi Photodetectors. \url{https://www.onsemi.com/products/sensors/photodetectors-sipm-spad}\relax
\mciteBstWouldAddEndPuncttrue
\mciteSetBstMidEndSepPunct{\mcitedefaultmidpunct}
{\mcitedefaultendpunct}{\mcitedefaultseppunct}\relax
\EndOfBibitem
\end{mcitethebibliography}

\end{document}